\documentclass[12pt]{article} \textheight=23cm \textwidth=16cm
\usepackage{amsmath} 
\usepackage{amsfonts}
\usepackage{amssymb} 
\usepackage{graphicx}

\begin{document}



\catcode`\@=11
\def\lesssim{\mathrel{\mathpalette\vereq<}}
\def\gtrsim{\mathrel{\mathpalette\vereq>}}
\def\vereq#1#2{\lower3pt\vbox{\baselineskip0pt \lineskip0pt
\ialign{$\m@th#1\hfill##\hfill$\crcr#2\crcr\sim\crcr}}}
\catcode`\@=12

\def\alt{\lesssim}
\def\agt{\gtrsim}
\bigskip

\begin{center}
{\Large \bf Connecting String Theory and Phenomenology\footnote{Invited
Plenary talk, SUSY02, Hamburg, June 2002, to appear in Proceedings,
ed. P. Nath and P. Zerwas}}

\bigskip

G.L.Kane

Michigan Center for Theoretical Physics

Randall Lab

University of Michigan

Ann Arbor, MI 48109
\end{center}

\begin{abstract}
To make progress in learning the underlying fundamental theory, it will
be necessary to combine bottom-up phenomenology and top-down analysis ---
in particular, top-down is unlikely to succeed alone. Here I elaborate on
the role of both, and describe obstacles that need to be overcome to
help data point toward the underlying theory, as well as approaches that
might help to bypass full systematic treatments. I also summarize
arguments that superpartners are probably being produced at the Tevatron
Collider.
\end{abstract}

\section{Introduction --- Top-Down and/or Bottom-Up?}

The central problem in particle physics continues to be understanding
the physics of electroweak symmetry breaking (EWSB).  Basically that
means finding and studying the Higgs boson(s).  Equally importantly we
need to confirm (or not) that weak scale supersymmetry is an essential
part of our low-scale world.  Then the central problems become
understanding the physics of supersymmetry breaking, and learning about
the underlying, more fundamental, theory at the high scale, where forces
unify.

I expect there will be little progress toward these goals without
experimental data.  Top down approaches are seldom successful in
physics.  One might think that theorists only need to write down string
theories, try out compactifications, get a 4D effective field theory
below the Planck scale, calculate predictions for low scale phenomena,
and soon have the basic answer.  Unfortunately, carrying out such a
process is much more difficult than it first seems to be.  Essentially
any prediction for data, or understanding of masses, requires knowing
$\tan\beta,$ the ratio of the two Higgs vevs.  But $\tan\beta$ does not
exist in the high scale theory.  It only arises as the EW symmetry is
broken, as one approaches the low scale.  $\tan\beta$ must be
calculated, and depends on many aspects of the high scale theory being
correct.  The origin of $\mu$ is equally important.  In my view a major
success of superstring theory is that $\mu=0$ occurs naturally in the
superpotential because $\mu$ enters as a mass term and the low-scale
theory is formulated from the string zero modes. Because of the
non-renormalization theorem this is stable.  But once supersymmetry is
broken a $\mu$-term will be generated, perhaps by the
Giudice-Masiero mechanism, or as a vev of a scalar field.  In the former
case the relevant contributions can occur as non-renormalizable
contributions in the Superpotential or Kahler potential, terms that may
not appear in lowest order and tree level. $\mu$ enters into many
predictions, for the Higgs sector, for chargino and neutralino and
sfermion masses and production cross sections and decay branching
ratios, and more.  If the origin of $\mu$ is not understood, a top-down
approach will not be fruitful. In a top down approach the two equations
from minimizing the Higgs potential and requiring it be bounded from
below would be used to calculate $m_Z$ and $tan\beta$ in terms of $\mu,
{m_H}_u, {m_H}_d$, and $b$, these four quantities all being determined
by the underlying theory.

One might think there are sectors for which predictions might be made,
such as the gluino mass, that do not depend on $\tan\beta$ or $\mu.$ But
it is more complicated.  Many parameters get connected by the needed RGE
running, which introduces dependence on squark masses, trilinear
couplings $A,$ and b (the soft parameter coefficient of $H_{u}H_{d}).$
$b$ is very important in determining $tan\beta$, and its RGEs depend
strongly on the trilinear couplings and gaugino masses. If any of these
are not known it is not possible to make useful low scale predictions.
Further, many of the soft parameters can be complex, and we now know
that their phases can have large effects on not only CP violating
phenomena but also on mass eigenvalues, production cross sections, decay
branching ratios, virtual effects, etc.  So the top-down approach must
also include an understanding of the phases and the origins of CP
violation, not only as a matter of principle but to do practical
calculations.  One might think that one could calculate charged lepton
masses.  The electron mass is too small to be meaningfully predicted ---
it is sensitive to many corrections from non-renormalizable operators
and loops.  Conceivably an approximate calculation of the ratio
$m_{\mu}/m_{\tau}$ could make sense, but it has some dependence on
non-renormalizable operators, $\tan\beta,$ trilinears, RGEs, etc.

So top-down approaches can and will be very important for guiding
thinking, but are unlikely to lead to detailed serious predictions that
really test the stringy ideas, even at the level of the 4D field theory,
since it will be necessary to make assumptions in order to get
predictions.  Then it is the assumptions that are being tested too (or mainly).

What about bottom-up?  It will be very exciting when superpartners are
found, and it will tell us low scale supersymmetry is indeed the right
direction for increased study, which knowledge is greatly needed.  But
it is crucial to realize that experimenters measure only information
about masses of mass eigenstates, from kinematics of particles that
enter the detector, and cross sections (times branching ratios).  Of all
the observables that can be measured, only one occurs in the Lagrangian
a theorist would write, and that only approximately (the gluino mass) .
The next stage is to study the pattern of superpartner properties to
learn about the underlying theory.  That will not be so simple ---
experimenters cannot communicate with string theorists, and vice versa,
because they do not share a common language.  Telling a string theorist
a chargino mass or cross section will be of no use --- those quantities
are not in the theory directly, and we saw above that it is very
unlikely there will be serious calculations for them.  Telling an
experimenter the high scale Lagrangian will not help with a comparison
of theory and data.

So it will be necessary to deduce the low-scale Lagrangian from the
low-scale data in order to connect experiment with theory.
Unfortunately, that cannot in general be done!  Every observable can be
expressed in terms of a set of complex Lagrangian parameters, such as
the gaugino masses, $\mu, $ squark masses, etc.  There is also always a
dependence on $\tan\beta.$ By simply counting one can see that at any
hadron collider the number of observables is always less than the number
of Lagrangian parameters, so the equations cannot be inverted to solve
for the parameters.  Indeed, this is the most important reason why we
need lepton colliders with polarized beams --- with polarization one
can double the number of measurements for non-zero cross sections, and
add additional observables by running at two different energies (since
the coefficients in the equations relating observables and Lagrangian
parameters depend on energy).

At the Tevatron, which is all we will have for at least five years from
this conference, the situation is worse because the Tevatron and its
detectors are luminosity-limited, even assuming Fermilab can get it to
work as well as was planned.  And by their very nature superpartners do
not give dramatic signals because every event has two escaping LSPs, so
no kinematical variable can show a dramatic peak of the sort that made W
and Z discovery relatively easy.  Indeed, that is a testable prediction
of supersymmetry. Instead, in several channels an excess of events will
slowly accumulate.  Probably no single signal will be statistically
significant.

Of course, the basic lesson of the introduction is that top-down and
bottom-up approaches are complementary, and must be merged to make progress.

\section{Learning the Low-Scale Theory}

In practice we may be lucky, and find that some parameters put us in a
region of parameter space where measurements are possible.  For example,
if $\tan\beta$ is very large it may be possible to observe
$B_{s}\rightarrow \mu\mu$ at the Tevatron and therefore get a
measurement of $\tan\beta$.  Data from the Higgs sector, the way the
electroweak symmetry is broken, how the hierarchy problem is solved,
gauge coupling unification, the absence of LEP signals, rare decays,
cold dark matter detectors, $g_{\mu}-2,$ proton decay, the neutrino
sector, and other non-collider physics will be very important to combine
with collider data to make progress.

The key point is that since supersymmetry is a real theory it is
possible to calculate its predictions for many processes and use them
all to constrain parameters.  Because of this even at hadron colliders
the situation may not be so bad.  By combining information from several
channels each with almost-significant excesses we can learn a lot about
the parameters and, more importantly, the basic theory.  For example,
the following table shows how different forms of supersymmetry breaking
and mediation lead to qualitatively different ``inclusive signatures''.

\smallskip
\hspace{-.4in}
\begin{tabular}{cccccccc}
Inclusive & $\widetilde{G}MSB,$ & $\widetilde{G}MSB,$ & $GMSB,$ &
Unstable & Gluino & $\underline{\left\langle D\right\rangle}$ \\
\underline{Signatures} & \underline{large $\mu$} & \underline{small
$\mu$} & \underline{low scale} & \underline{LSP} &
\underline{Condensation} & \\
Large E$_{T}$ & yes & yes & yes & no &  &  & \\
Prompt $\gamma^{\prime}s$ & no & sometimes & yes (but...) & no &  &  & \\
Trilepton events & yes & no & no & no &  &  & \\
Same-sign dileptons &  &  &  &  &  &  & \\
Long-lived LSP &  &  &  &  &  &  & \\
$\tau-$ rich &  &  &  &  &  &  & \\
$b-$ rich &  &  &  &  &  &  &
\end{tabular}

$\bigskip$

One can add both rows and columns --- this is work in progress.  This
approach also shows how to combine top-down and bottom up approaches ---
one uses top-down analysis to identify the columns and fill in the
missing entries in the table.  By simply identifying qualitative
features of the channels with excesses one can focus on a few or even
one type of theory.  Then detailed study can let one zoom in on the
detailed structure of the underlying theory and its high energy
features.  With such an approach one can partly bypass the problem of
not being able to fully isolate the Lagrangian explicitly.  One will not
be able to prove that specific superpartners are being observed with
this ``inclusive'' analysis, but we can gamble and leave the proof for
later.  In this table $\widetilde{G}$ stands for gravitino, and
$\widetilde{G}MSB$ for gravity-mediated supersymmetry breaking, $GMSB$
for gauge-mediated supersymmetry breaking, $\left\langle D\right\rangle
$ for supersymmetry breaking by an D-term vev, etc.  The key point is
that each inclusive observation allows one to ``carve'' away part of the
parameter space, and the remaining parts point toward the underlying
high scale theory.  One does not need to measure every soft parameter to
make progress, because the patterns, the mass orderings, etc., imply
much about the underlying theory --- if one understands the theory.

Can we make progress while we know so little about string theory and how
to find its vacuum state, and with very limited data?  Perhaps history
can provide guidance here.  At the end of the 1960s particle physics was
widely viewed as being in bad shape, with little hope for significant
progress.  Three years later the Standard Model existed, and most active
workers were convinced it was correct.  The progress was based on some
seemingly isolated theoretical results, including Yang-Mills gauge
theories, the Higgs mechanism, the Glashow electroweak model and the
Weinberg lepton model, plus clues from the hadron spectrum.  The
experimental results were the knowledge that weak interactions occurred
via V,A currents (rather than S,P,T), parity and charge conjugation
violation (i.e. chiral fermions), knowing the weak interactions were
weak, and the SLAC deep inelastic scattering data that implied
constituents in the proton.  It all fell into place with the proof of
the renormalizability of the electroweak theory and asymptotic freedom.
We may be in a similar situation once the existence of Higgs bosons and
light superpartners is confirmed at the Tevatron. Sometimes people say
we know too little about string theory to try to do phenomenology with
it. I think it is the opposite --- only if we learn from phenomenology
what region of string theory space to focus on are we likely to make
progress.

\section{Obstacles}

But there are a number of obstacles that can obscure the connection
between low-scale data and low-scale theory, and the connection to the
high-scale theory.  These obstacles can be at least partly overcome by
theoretical study --- this is a fruitful area for research, particularly
now that we are close to getting the data.  Here I will list a number of
obstacles and sometimes comment on how they might be studied.

\begin{itemize}
\item Most theories have intermediate scale matter\cite{a, b} that can
affect RGE running even though the theory remains perturbative up to the
high scale.  We have to learn how to find ways to run up that are not
sensitive to the intermediate scale matter, or find checks such as
running two different ways that would give the same answer without
intermediate scale matter but differ in its presence, or find other ways
to either learn about such particles or bypass them. For example, some
types of intermediate scale matter increase the fine-tuning needed to
explain EWSB, so they can be excluded.

\item We don't know at what scales to start or end RGE running.  In
general the supersymmetry breaking scale will be different from the
mediation scale.  In string theories in general the string scale is
different from the unification scale or the compactification scale.
Certainly the literature is not consistent on this issue.  Model studies
should be carried out that clarify how to proceed in practice.
Phenomenological results are certainly sensitive to these scales. In
general we can expect patterns to emerge that will help settle such
questions.

\item Infrared fixed points\cite{c, d} would make it difficult to deduce
  high-scale quantities.  This occurs when a range of high-scale input
  parameters flow to the same low-scale value, so a measurement of the
  low scale value implies a range of high-scale values.  An important
  conclusion\cite{e} is that accurate low-scale measurements, such as
  can be obtained at linear colliders, will be important for deducing
  the high-scale theory.  Further theoretical study is needed to learn
  how to combine measurements to avoid the ambiguities. The philosophy
  of fixed points is interesting. Some people view them as a good thing
  because one can predict or understand a low scale result even if we do
  not know the high scale theory. But actually it is the opposite --- we
  want to learn the underlying high scale theory, so we need to get
  around fixed point behavior.

\item Most soft terms can be complex\cite{f, g}.  While there are
constraints on their phases, suggesting some may be small, the phases
may affect many observables such as the Higgs sector, superpartner cross
sections, etc.  If they are present but not included in analysis wrong
conclusions will be drawn.  EDM data, and the apparent success of the SM
description of CP violation in the K and B systems, suggest the soft
phases are small, but no known symmetry or principle implies that the
soft phases should be small, so care is needed in how phases are
included (or not included).

\item Additional U(1) symmetries under which visible sector particles
are charged may lead to D-terms\cite{h} that affect scalar masses even
though the U(1) symmetry is broken at a high scale.  That can shift
sfermions, and in particular $M_{H_{u}}^{2}$ and $M_{H_{d}}^{2},$ thus
changing the way EWSB works, and therefore leading to changes in $\mu,$
gaugino masses, CDM relic density and detectability, etc.  Conversely,
from the observed pattern of scalar masses once superpartners are
observed it may be possible to deduce their U(1) charges and learn about
the high-scale U(1)s.

\item The theory may have an extended gauge group, i.e. a grand
unification, and/or extra U(1)s.  This can also lead to extra soft
terms, larger mass matrices for neutralinos and for the Higgs sector,
new gauge couplings, kinetic mixing, etc.  If one does not include the
larger gauge group in the analysis (because one does not know about it)
the result may be misinterpretation of low-scale data and its
implications. For example, the trace of the neutralino mass matrix is
the sum of its eigenvalues. If it is 5x5 instead of 4x4, with an extra
soft term $M_{1}^{\prime},$ then the true trace would be
$M_{1}^{\prime}+M_{1}+M_{2}$ but this would be interpreted as just
$M_{1}+M_{2},$ so one would deduce incorrect values for $M_{1}$ and
perhaps for $M_{2}.$ One can develop consistency checks to detect such
departures from the simplest theories.

\item High scale threshold corrections must be made, but of course
cannot be made until the high scale spectrum is known.  So their effects
must be included in a way that allows them to be determined and does not
lead to wrong conclusions about the implications of low energy
phenomena.

\item In general there will be non-renormalizable operators that affect
small Yukawa couplings and therefore the CKM phase, calculation of
$\sin^{2} \theta_{W}$ and other precision data, etc.  If such operators
are present it will be necessary to rescale gauge couplings and gaugino
masses and even the ratio $M_{a}/g_{a}^{2},$ if the gauge kinetic
function and the Kahler potential have such operators.

\item The Kahler potential can have flavor-dependent contributions that
affect Yukawa couplings and thus the CKM angles and phase, and trilinear
couplings, and thus the analysis of rare decays and lepton
flavor-violating effects.
\end{itemize}

More obstacles can be listed.  Perhaps it is best to think of them not
as obstacles but as opportunities for research.  By proceeding with a
judicious combination of study of patterns in data and clues, plus
models and top-down analysis, it may be possible to deal with the
obstacles so they do not prevent progress toward learning the underlying
theory.

\smallskip

\section{Superpartners}

If we are sure that low energy supersymmetry is part of the description
of nature, the absence of effects in some processes contains a huge
amount of information.  So once direct evidence for the light Higgs
sector and a few superpartners is obtained, we can draw many conclusions
about the form of supersymmetry breaking and mediation, and the
structure of the underlying theory and its stringy nature, not only from
what is explicitly seen but also what is absent.  Where are the
superpartners?  Is there reason to expect them to be found
experimentally?  This is of course an old question, which we revisit.
We are motivated by the absence of superpartners at LEP on the one hand,
and by increased theoretical understanding of supersymmetry theory on
the other.

Should superpartners have been observed at LEP, or in Run I at the
Tevatron?  What emerges from study of modern models is that
superpartners might have reasonably been in the LEP or Run I parameter
region, but we would have been lucky if they were, in the sense that
much of the reasonable parameter space puts the superpartners somewhat
above those regions --- for LEP it is the energy cutoff, while for the
Tevatron it is the luminosity that was and is limited.

What do we know that can set a mass scale for superpartner masses?  From
a purely theoretical side we would have to understand supersymmetry
breaking to set that scale, and our knowledge of the mass spectrum that
would result from supersymmetry breaking is far from implying any
particular mass scale for superpartners.  We must use some physics
input.  We could of course just ignore the question and wait, but our
view of what masses to expect can significantly affect what resources
are available for the search, and what methods are used.

What physics do we know that is relevant?  First, the value of
$m_{h}^{2}$ is sensitive to the scale of new physics and essentially
rises to that scale.  Qualitatively this is the hierarchy problem, and
tells us that the new physics, presumably the superpartner masses, are
``around a TeV'', rather than tens of TeV or much higher, but is not
quantitative enough to distinguish a TeV from 100 GeV.

Second, gauge coupling unification implies light superpartner masses and
small $\mu.$ The latter is sometimes forgotten, but $\mu$ enters
chargino and neutralino and squark and Higgs masses, so it significantly
affects the gauge coupling running and the location of the thresholds
where the supersymmetry beta functions enter the running.  But again the
results are not quantitative enough to pin down the superpartner masses
within an order of magnitude. If the LSP is indeed the CDM then getting
the right relic density provides constraints on the soft parameters, but
there is a large range of parameters that give the right relic density,
and we do not know how much of the relic density is the LSP.

It turns out that the only thing we know that quantitatively relates the
soft masses to a measured number is radiative electroweak symmetry
breaking, REWSB.  \textit{IF} supersymmetry indeed provides the
explanation for how the electroweak symmetry is broken, to allow fermion
and gauge boson masses, then it provides an equation relating $M_{Z}$ to
soft masses,

\begin{eqnarray}
M_{Z}^{2}=-2\mu^{2}+2\left(  \frac{m_{H_{d}}^{2}-m_{H_{u}}^{2}\tan^{2}\beta
}{\tan^{2}\beta-1}\right)
\end{eqnarray}

\noindent at tree level.  This is the only presently known connection of
the supersymmetry parameters to data that is precise enough to be
useful, and is extremely important, though not yet well understood.  One
can rewrite this in a useful form by expressing the quantities on the
right hand side, which are here evaluated at the weak scale, in terms of
high scale input.  We do that in the general MSSM, with the full soft
Lagrangian and no assumptions about gaugino or squark mass degeneracies,
obtaining\cite{l}

\begin{eqnarray}
M_{Z}^{2}=-1.9\mu^{2}+6.9M_{3}^{2}-0.3M_{2}^{2}+0.01M_{1}^{2}-1.2m_{H_{u}}
^{2}+1.6M_{Q}^{2}+...
\end{eqnarray}

\noindent Here the soft masses (and $\mu)$ are evaluated at the
unification scale. This equation is written for $\tan\beta=5.$

\emph{IF }supersymmetry does indeed explain the origin of $M_{Z},$ we
would naively expect none of the terms on the right hand side to be much
larger than $M_{Z}^{2}.$ Accidental cancellations do not explain basic
results in physics.  That implies upper limits on $M_{3},$ $\mu,$ etc.
The limits are a little soft, but it turns out that because the
coefficients are large even letting each term be a few $M_{Z}^{2}$ still
leads to important upper limits on the superpartner masses.  In the past
people have gone through such reasoning and deduced that very light
superpartners should exist.  When they were not found at LEP people got
confused about the arguments.  But the past arguments were based on
using assumed degeneracies.  If the three gaugino masses are taken
equal, $M_{3}=M_{2}=M_{1}=M_{1/2},$ then the first three terms combine
into 6.6 ${M^2}_{1/2}$ so that $M_{1/2}$ is expected to be quite small,
implying very light charginos and neutralinos that do not exist.  Here
we see that without the degeneracy assumption there are essentially no
restrictions on $M_{1}$ and $M_{2},$ so the absence of charginos and
neutralinos at LEP should not be thought to be evidence against the
argument based on REWSB, but rather evidence against gaugino mass
degeneracy.

Sometimes people think gaugino mass degeneracy is implied by gauge
coupling unification.  But that is not so.  Our best theoretical
guide\cite{b, i} to such questions is to ask what happens in string
theory.  While tree level gaugino masses are often degenerate, they are
often suppressed in string theories (see\cite{i} and references therein).
The one-loop corrections are typically not degenerate and then lead to
non-degenerate gaugino masses.  At the same time, the gauge coupling
leading contributions are not suppressed, so loop effects are small, and
the leading contributions unify.  More theoretically, $M_{a} $ and
$g_{a}^{2}$ arise from vevs of different components of the dilaton
multiplet, so there is no general connection at all.  Sometimes people
argue that the RGE invariance of $M_{a}/g_{a}^{2}$ implies gaugino mass
degeneracy, but again it only holds at tree level.  Phenomenologically
one can make a strong statement.  As remarked above, gaugino mass
degeneracy plus some constraints on fine tuning imply that $M_{1}$ and
$M_{2}$ are light enough to be inconsistent with the absence of
charginos and neutralinos at LEP.  Thus gaugino mass degeneracy requires
large fine tuning.

Are there possible loopholes that would allow one to evade the
conclusion from equation 2 that $M_{3}$ and $\mu$ must be at most about
$M_{Z}$ in size?  We\cite{b} have recently pursued this question, both
by examining models and by looking at how stringy approaches behave.  It
seems sensible to make stringy models by making plausible assumptions
when needed, and examine how they explain EW symmetry breaking.
Basically we find that $\mu$ and the various soft masses arise in such
different ways physically that cancellations are extremely unlikely even
in very general frameworks.  One can see that the variation with
$m_{top}$ is small, and the variation with $\tan\beta$ is mild for
$\tan\beta\agt 4$.  There are essentially four possibilities to discuss.

\begin{enumerate}
\item Could $\mu$ and $M_{3}$ be related so that their contributions
cancel in the calculation of $M_{Z}?$ Answering that requires an
understanding of how $\mu$ arises.  One of the phenomenological
successes of supersymmetric string theory is in setting $\mu$ to zero
naturally at the unification scale.  Then $\mu$ can arise by a
Giudice-Masiero like mechanism, from terms in the Kahler potential or
non-renormalizable terms in the superpotential, or alternatively $\mu$
can arise from NMSSM type models where a scalar gets a vev, or stringy
generalizations where scalars in non-renormalizable operators get vevs.
Either way if one writes examples one sees that $\mu$ depends on very
different physics from $M_{3}$.  Particularly when one looks in examples
with gauge coupling unification one finds that it is extremely hard to
imagine any robust cancellation between $M_{3}$ and $\mu.$

\item Could there be cancellations among soft terms, either among
gaugino masses or between gaugino masses and scalars?  A cancellation
among gaugino masses would require a huge non-degeneracy since the sizes
of their contributions to $M_{Z}$ are so different.  If the sizes are
made less different by reducing $M_{3}$ that would give a light gluino
as needed.  What about the scalars and $M_{3}$?  Again, that would
require a large scalar non-degeneracy.  More generally, one can see that
the soft parameters are affected by supersymmetry breaking and by how
the breaking is transmitted, introducing at least two mass scales.  Or
one can think of it in terms of dilaton dynamics (which strongly affects
gaugino masses) and moduli dynamics (which strongly affects scalar
masses).  To have a significant cancellation would require special
relations among these separate parts of the theory.  If one thinks in
terms of parameters, there would have to be a special relation among
$m_{3/2},$ dilaton vevs, and moduli vevs.

\item Can the coefficients be made smaller?  Yes.  For example, one can
run down from a lower scale, in which case the coefficient of $M_{3}$ is
smaller.  Or include various amounts of intermediate scale matter in
such a way as to reduce the coefficients.  But first, the coefficients
do not get small, just smaller.  More important, examining any
particular approach that reduces the coefficients shows that the same
physics also leads to lighter superpartner masses.  For example, in the
MSSM run down from the usual unification scale one finds the physical
gluino mass $m_{\tilde{g}} \approx3M_{3}^{UNIF},$ while if one reduces
the coefficient of $M_{3}$ from 6.9 by a factor of 10, allowing a larger
$M_{3}$ for a given amount of fine tuning, the physical gluino mass in
the reduced coefficient case turns out to be
$m_{\tilde{g}}\approx0.4M_{3}^{UNIF}.$ Thus the physical gluino is again
light.

\item Another argument that superpartners are heavy compared to what is
detectable at the Tevatron, or even at a 500 GeV linear collider, has
emerged with the construction of so-called benchmark models in recent
years.  For example, only one of the widely publicized Snowmass
models\cite{j} has superpartners light enough so they may possibly be
detectable at the Tevatron.  If such a result is generic it would be
important.  Fortunately, it is not generic.  We\cite{i} have constructed
some models that are well-motivated theoretically, based on the
heterotic string.  Of course one cannot yet derive the needed
compactifications or deduce how supersymmetry is broken, but one can
assume plausible mechanisms that could happen, and specialize to
concrete models.  We find sets of models that have less fine tuning than
the minimal SUGRA type Snowmass or CERN\cite{k} models, and in
particular have light enough gauginos so that some can always be
detected at the Tevatron (and at a 500 GeV linear collider).  The bottom
line is that the concrete models that have been previously constructed
are, for historical reasons, a very special set.  They can only
accommodate EWSB by having large cancellations, which casts doubt on
their relevance.  They may also be misleading for their implications for
collider signatures and background studies.  They do not imply that
nature's superpartners are heavy compared to what can be seen at the
Tevatron.
\end{enumerate}

\section{Summary}

From the points made here we can conclude that if superpartners are not
produced at the Tevatron it is rather unlikely that supersymmetry and
the usual radiative mechanism is the actual explanation for electroweak
symmetry breaking.  Then the constraints on superpartner masses are very
weak.  Gauge coupling unification depends on the same conditions as
radiative EWSB, requiring $\mu$ and soft masses at the TeV scale, though
it depends on the masses logarithmically rather than quadratically.
Nevertheless, if superpartners are not produced at the Tevatron it is
interesting to consider what arguments imply they should be seen at LHC.

If superpartners are observed at the Tevatron as should be expected,
then considerable analysis and good thinking will be needed to deduce
the Lagrangian at the weak scale from the measured masses and cross
sections and branching ratios.  Further innovative thinking and analysis
will be needed to learn the unification scale Lagrangian, and to deduce
properties of supersymmetry breaking and mediation from the low scale
information.  In both cases there are important research
opportunities. Combining inclusive analysis of collider signals with
information from rare decays, dark matter, neutrino physics, proton
decay, baryogensis, dipole moments and more may allow us to bypass many
apparent obstacles. One can imagine that it will be possible to guess
the supersymmetric Standard Model from the resulting analysis, then the
4D effective string field theory, and perhaps even the 10D string
theory.

\section{Acknowledgments}

I am grateful to my collaborators, and particularly to Lisa Everett,
Lian-Tao Wang, James Wells, and Brent Nelson, for discussions and
lessons. I want to thank the DESY Theory Group, and particularly Peter
Zerwas, for their good hospitality.

\end{document}